# Validation and noise robustness assessment of microscopic anisotropy estimation with clinically feasible double diffusion encoding MRI

*Leevi Kerkelä[1], Rafael Neto Henriques[2], Matt G. Hall[1,3], Chris A. Clark[1], Noam Shemesh[2]*

1) UCL Great Ormond Street Institute of Child Health, University College London, London, United Kingdom
2) Champalimaud Neuroscience Programme, Champalimaud Research, Champalimaud Centre for the Unknown, Lisbon, Portugal
3) National Physical Laboratory, Teddington, UK

**Correspondence**
Dr. Noam Shemesh, Champalimaud Neuroscience Programme, Champalimaud Centre for the Unknown, Av. Brasilia 1400-038, Lisbon, Portugal.
Email: noam.shemesh@neuro.fchampalimaud.org

Word count: 5066.

# ABSTRACT

**Purpose:** Double diffusion encoding (DDE) MRI enables the estimation of microscopic diffusion anisotropy, yielding valuable information on tissue microstructure. A recent study proposed that the acquisition of rotationally invariant DDE metrics, typically obtained using a spherical "5-design", could be greatly simplified by assuming Gaussian diffusion, facilitating reduced acquisition times that are more compatible with clinical settings. Here, we aim to validate the new minimal acquisition scheme against the standard DDE 5-design, and to quantify the proposed method's noise robustness to facilitate future clinical use.

**Methods:** DDE MRI experiments were performed on both *ex vivo* and *in vivo* rat brains at 9.4 T using the 5-design and the proposed minimal design and taking into account the difference in the number of acquisitions. The ensuing microscopic fractional anisotropy (µFA) maps were compared over a range of *b*-values up to 5000 s/mm$_2$. Noise robustness was studied using analytical calculations and numerical simulations.

**Results:** The minimal protocol quantified µFA at an accuracy comparable to the estimates obtained via the more theoretically robust DDE 5-design. µFA's sensitivity to noise was found to strongly depend on compartment anisotropy and tensor magnitude in a non-linear fashion. When µFA < 0.75 or when mean diffusivity is particularly low, very high signal to noise ratio (SNR) is required for precise quantification of µFA.

**Conclusion:** Our work supports using DDE for quantifying microscopic diffusion anisotropy in clinical settings but raises hitherto overlooked precision issues when measuring µFA with DDE and typical clinical SNR.

# INTRODUCTION

Diffusion-weighted MRI (dMRI) is sensitive to micron-scale displacements, making it a useful method for measuring microstructural features[1]. Diffusion tensor imaging (DTI) represents diffusion in a voxel by a single Gaussian diffusion tensor with characteristic size, shape, and orientation[2]. At low diffusion weighting, even in the case of restricted diffusion, dMRI acquisition schemes with at least six non-collinear directions produce approximately mono-exponential signal decays from which the diffusion tensor's eigenvectors and eigenvalues can be quantified and other quantitative metrics such as mean diffusivity (MD), fractional anisotropy (FA), and absolute orientation are typically derived[2]. DTI has found widespread application in both clinical research and medicine[3–6], where it is regularly used in studying white matter tissues[7,8], acute stroke characterization[9,10], and pre-surgical planning[11], for example. However, the single tensor representation is not always sufficient for characterizing complex systems: restricted diffusion or polydispersity in diffusion components can induce non-mono-exponential signal decays, which cannot be captured by the DTI representation. Diffusion kurtosis imaging[12] (DKI) methods have been developed to extend the signal representation[13] to the second order in $b$, while $q$-space imaging methods[14–16] exploit the Fourier relationship between the averaged propagator and diffusion-weighted signal to enhance microstructural sensitivity.

The methods mentioned above belong to a class of single diffusion encoding (SDE)[17] methods because they probe the displacements of spins along a single axis. Despite their usefulness in characterizing tissues, a fundamental limitation of SDE methods is their inherent averaging over the ensemble of diffusion compartments in the imaging voxel. For example, in a system comprising an ensemble of diffusion tensors, all with identical magnitude and anisotropy but their directors aligned according to an orientation distribution function, SDE-driven anisotropy will typically be lower than the true underlying microscopic anisotropy (µA) – the anisotropy of the individual tensor in its own frame of reference[17]. In the extreme case of powder-averaged systems, SDE's anisotropy will approach zero even if the underlying µA is very high. Microscopic fractional anisotropy (µFA) is another important metric, representing the normalized microscopic anisotropy, which is independent of the magnitude of the diffusion tensor[17–19].

Multidimensional diffusion encoding (MDE) methods[20], such as double diffusion encoding (DDE) and *q*-space trajectory encoding[21,22] (QTE), have gained significant attention over the previous years for their ability to disentangle microscopic diffusion anisotropy and orientation coherence[19,23,24], potentially capturing clinically-relevant information[25,26] about tissue microstructure that is inherently unavailable in traditional SDE experiments. While DDE in the long mixing time and long diffusion time regime is a model-free method for measuring µA[19,27,28], time-independent Gaussian diffusion is assumed when measuring µA with QTE[20,22].

In the context of microstructural estimation, DDE was first proposed by Cory et al. in 1990 for measuring diffusion correlations along different dimensions in powder averaged systems[29], thereby reporting on µA. The DDE methodology has been reviewed in the past[17,30]; briefly, DDE sequences consist of two diffusion-sensitizing time periods spanned by two diffusion-encoding gradient pulse pairs, separated by a mixing time (Figure 1A). Mitra predicted that in long mixing time angular DDE experiments, the amplitude of the signal modulation as a function of the angle between the two gradient pulse pairs is sensitive to µA in powder averaged systems[31]. In 2002, Callaghan and Komlosh detected microscopic anisotropy in polydomain lyotropic liquid crystal systems by combining parallel and orthogonal gradient pulse pairs[14]. While several DDE studies later obtained microscopic anisotropy in powder averaged systems ranging from porous media[32] to biological tissues[33], Lawrenz and Finsterbusch were the first to propose a rotationally invariant acquisition scheme for extracting µA[34]. Jespersen et al. proposed a more robust rotationally invariant acquisition methodology termed the DDE 5-design, which is a set of wave vector rotations that estimates the average over all possible wave vector rotations[19], requiring 72 unique acquisitions. Bias, as well as correction schemes, have been recently reported for such acquisitions due to higher order terms[35,36], yet they require multiple diffusion-weighting shells, prolonging scan time. Along with poor vendor availability, the large number of acquisitions required for obtaining a theoretically justified powder average has likely impeded DDE from becoming more common in clinical settings.

Recently, Yang et al. proposed that, if diffusion can be assumed to be Gaussian within microscopic domains, as few as 12 acquisitions could be sufficient for measuring µA and µFA[26] (hereafter defined as the "minimal" design). This simplification is based on the idea that if diffusion in every microscopic compartment can be fully characterized by a diffusion tensor, some acquisitions with orthogonal wave vectors in the 5-design are redundant as they correspond to the same in-plane trace of the diffusion tensor. Additionally, asymmetric sampling of diffusion-weighting directions covering half a sphere is sufficient in the absence of bulk flow. This realization allows the 5-design to be simplified from 12 + 60 acquisitions to 6 + 15 acquisitions for parallel and orthogonal wave vectors, respectively. Using simulations of infinite cylinders with orientation dispersion and *in vivo* imaging experiments with a clinical whole-body scanner, Yang et al. experimentally showed that the design can be further reduced down to 6 + 6 without significant differences in the observed value of microscopic diffusion anisotropy. Importantly, such a dramatic reduction in directions, and hence in acquisition time, have strong implications for clinical applications and indeed, Yang et al. measured µFA maps in the clinical setting using the minimal design[26]. However, to our knowledge, this new minimal approach has not been directly compared to the more theoretically robust DDE 5-design. Clearly, validation is important, because new methods may produce biased results[37,38], and bias due to an insufficient number of measurement directions in powder averaged measurements acquired with MDE has been reported[39]. Therefore, in this study, we sought to experimentally validate the proposed approach by directly comparing its extracted metrics to their 5-design counterparts over a range of relevant *b*-values. Additionally, we sought to assess the *precision* of μFA estimates derived from DDE measurements, which have been scarcely explored previously. We studied the noise robustness of μFA derived from the clinically feasible DDE minimal design in detail using analytical calculations and noise simulations in order to provide guidelines for its use in neuroscientific and clinical research.

# THEORY

## µFA measurements

In arbitrarily organized systems in terms of orientation coherence, obtaining a powder averaged signal $S_{iso}$ would entail sampling over all possible orientations of the gradient pulse pair directions:

$$S^{iso}(q,\theta) = \frac{1}{8\pi^2} \int_{SO(3)} S(Lq\hat{e}_1, Lq\hat{e}_2) \, dL \tag{1}$$

where $S$ is the measured signal, $q$ is the magnitude of the diffusion wave vector, $\theta$ is the angle between the pulse pair directions represented by unit vectors $\hat{e}_1$ and $\hat{e}_2$, $L$ represents a rotation, and $SO(3)$ is the rotation group. It was shown by Jespersen et al. that if $S$ is equal to the signal cumulant expansion up to the fifth order in $q$, the integral in equation 1 can be calculated exactly with a finite number of rotations[19]:

$$S^{iso}(q,\theta) \approx \frac{1}{|\chi|} \sum_{L \in \chi} S(Lq\hat{e}_1, Lq\hat{e}_2) \tag{2}$$

The set of wave vector rotations $\chi$ is known as the 5-design which consists of 12 rotations for parallel gradient pairs and 60 rotations for non-parallel (typically, orthogonal) gradient pairs (Figure 1B). Yang et al. recently proposed that, by assuming Gaussian diffusion and orientation dispersion, the 5-design can be reduced to a minimum of only six parallel and six orthogonal gradient pulse pair rotations[26].

Given the powder averaged data acquired with parallel and orthogonal pulse pairs, microscopic anisotropy (µA) can be estimated as[19]

$$\varepsilon = (\mu A)^2 = ln\left(\frac{S_\parallel^{PA}}{S_\perp^{PA}}\right) b^{-2} \tag{3}$$

where ∥ and ⊥ stand for acquisitions with parallel and orthogonal wave vectors, *PA* stands for powder average, and *b* is the *b*-value of one of the gradient pulse pairs in the DDE sequence. µA is a measure of the average variance of the eigenvalues of the microscopic diffusion tensors, irrespective of orientation, and as such it depends on the microscopic tensors' anisotropy *and* size[19,40]. Microscopic fractional anisotropy (µFA) is a normalized measure of microscopic diffusion anisotropy that only depends on anisotropy of the diffusion tensors.

$$\mu FA = \sqrt{\frac{3}{2}} \sqrt{\frac{\varepsilon}{\varepsilon + \frac{3}{5}(MD)^2}} \qquad (4)$$

where MD stands for mean diffusivity[19]. In realistic voxels, a distribution of microscopic diffusion orientations will nearly invariably exist, and thus the measured µA and µFA will correspond to the average eigenvalue variance[38].

**Accuracy of µFA**

Since the formalism presented above is based on the truncation of the signal cumulant expansion, µA estimates can be corrupted by non-vanishing higher order effects. Accuracy problems arising due to these reasons were extensively discussed by Ianus et al. who proposed that the accuracy of µFA can be improved by introducing a higher order correction ($P_3$) to equation 3 [35]:

$$\ln\left(\frac{S_\parallel^{PA}}{S_\perp^{PA}}\right) = (\mu A)^2 b^2 + P_3 b^3 \qquad (5)$$

This correction allows µA and µFA to be more accurately measured.

**Precision of µFA**

The higher order corrections result in a more accurate estimate of µFA, albeit at some cost in precision due to the third order polynomial fit being more sensitive to noise than the second order fit. To our knowledge, previous studies have not addressed how µFA's noise robustness is affected by ε's logarithmic dependence on the ratio of the powder averaged parallel and orthogonal wave vector acquisitions (Equation 3), the normalization of ε by MD (Equation 4) and the ensuing non-linear dependence on ε and MD. This can be examined using a common formula for propagation of uncertainty[41], that approximates the error in $f$, which depends on vector **x** containing noisy measurements, due to random noise as

$$\sigma_{f(x)}^2 = \sum_{i=1}^{N} \sigma_i^2 \left(\frac{\partial f}{\partial x_i}\right)^2 \tag{6}$$

where $\sigma_{f(x)}$ represents the standard deviation of noisy $f$, and $\sigma_i$ is the standard deviation of noisy measurements of $x_i$ [41]. This first-order approximation is valid if the errors are statistically independent and small compared to the signal level[41]. Applying Equation 6 on Equations 3 and 4 gives

$$\sigma_\varepsilon = \sqrt{\frac{\sigma^2}{12 S_\parallel^2} + \frac{\sigma^2}{12 S_\perp^2}} b^{-2} \tag{7}$$

$$\sigma_{\mu FA} = \sigma_\varepsilon \sqrt{\frac{27}{200}} \frac{(MD)^2}{\sqrt{\varepsilon}(\varepsilon + \frac{3}{5}(MD)^2)^{\frac{3}{2}}} \tag{8}$$

where it was assumed that the powder averaging is done over 12 directions for both parallel and orthogonal acquisitions (Figure 1C), and that MD can be observed without error. This assumption was made because MD can be estimated from the first order term of the signal cumulant expansion, making it more robust to noise than µFA, which is estimated from the second order term[13]. It can be seen from Equations 7 and 8 that the magnitude with which error propagates to µFA depends on both $\varepsilon$ and MD in a non-linear way, and that the same level of noise results in a larger uncertainty in µFA when $\varepsilon$ or MD are very low, which in turn impedes precise measurements of µFA in such voxels. On the contrary, high MD will result in very diminished signal, effectively reducing SNR and preventing precise measurements due to the noise floor.

# METHODS

All animal experiments were preapproved by the competent institutional and national authorities, and carried out according to European Directive 2010/63.

**Specimen preparation**

*Ex vivo experiments:* A rat brain (N = 1) was extracted through standard transcardial perfusion from a healthy adult animal and was then immersed in 4 % paraformaldehyde (PFA) solution for 24 h, followed by immersion in a phosphate-buffered saline (PBS) solution for at least 48 h. The extracted brain was inserted into a Fluorinert (Sigma Aldrich, Lisbon, PT) filled 10-mm NMR tube and placed in the scanner at 23 °C.

*In vivo experiments:* A healthy rat (N = 1, age 135 days, weight 288 g) was anesthetized with isoflurane (4% for induction, 2% maintenance delivered via nose cone) and placed in the scanner. Respiration rate and rectal temperature were monitored and kept stable over the entire experiment via small adjustments in isoflurane levels and circulating warm water, respectively.

**MRI experiments**

All experiments were performed on a 9.4 T Bruker (Karlsruhe, Germany) BioSpec scanner harnessing an 86 mm volume coil for transmission and 4-element array cryocoil for reception. A DDE-EPI pulse sequence written in-house was used with the following experimental parameters: $\delta$ = 5 ms and $\Delta = \tau$ (mixing time) = 15 ms, TE = 69 ms in experiment 1 and TE = 65 ms in experiments 2 and 3, TR = 1 s, FOV = 20 mm x 20 mm, matrix of 100 x 100 (partial Fourier factor = 1.25), leading to an in-plane resolution of 0.2×0.2 mm2, 3 slices in experiment 1 and 5 slices in experiments 2 and 3 of 0.8 mm thickness were acquired with a slice gap of 0.5 mm. The echoes were double-sampled to suppress residual ghosting.

The accuracy of the proposed minimal design[26] in neural tissue was validated using the following three experiments.

**Experiment 1:** To avoid any issues associated with *in vivo* measurements, an *ex vivo* rat brain was scanned with both the minimal design and the 5-design using *b*-values typically used in a clinical setting, namely, three b-values equally distributed between 1000 and 3000 s/mm$^2$. To maximize SNR, the 5-design measurements were averaged over 30 acquisitions and the minimal design measurements over 90 acquisitions (the factor of 3 in number of averages accounts for multiplexing). The 5-design experiment here was repeated twice to measure the inherent variance in μFA due to finite SNR.

**Experiment 2:** In addition to experiment 1, the *b*-value dependency of our comparison was investigated by imaging the *ex vivo* brain with more densely sampled and higher b-values, specifically, using ten *b*-values evenly distributed between 500 and 5000 s/mm$^2$. The 5-design measurements were averaged over 8 acquisitions and the minimal design measurements over 24 acquisitions, accounting for multiplexing.

**Experiment 3:** To avoid bias due to the fixation process and temperature, the comparison of the two methods was repeated *in vivo*. Here, the experiment was performed with *b*-values of 1000 and 2500 s/mm$^2$. The 5-design measurements were averaged over 8 acquisitions and the minimal design measurements over 24 acquisitions, again accounting for multiplexing.

The *b*-values expressed here refer to the total diffusion-weighting during the DDE experiment. For every 5-design acquisition, 13 b = 0 s/mm$^2$ images were acquired for temporal SNR estimation. In all experiments involving the minimal design, rather than using 6 parallel and 6 orthogonal wave vector pair directions distributed over half a sphere as originally prescribed[26], 12 + 12 directions (Figure 1C) distributed over the surface of the sphere were used to eliminate the possibility of artifacts arising from cross-terms between the imaging and diffusion gradients[42]. The 5-design was acquired as usual with 12 parallel and 60 orthogonal directions (Figure 1B).

**Image analysis**

**Preprocessing:** Diffusion-weighted data was denoised using a Marchenko-Pastur-PCA denoising procedure with a 17 x 17 x 1 sliding window[43]. Gibbs ringing artifacts were reduced using a sub-voxel shift algorithm[44]. To correct for motion and signal drifts, data was registered to the first non-diffusion-weighted image with a sub-voxel discrete Fourier transform algorithm[45]. No procedure for eddy current induced artifact correction was applied, since no eddy current distortions could be visually observed in any of the datasets.

**Analysis:** µFA maps were calculated from the powder averaged data without the higher order correction using Equations 3 and 4. MD used in calculating µFA was estimated by fitting a diffusion tensor to the data acquired with parallel wave vectors at $b = 1000$ s/mm$^2$ using DiPy[46]. Voxel-specific temporal SNR was quantified as the mean signal divided by the standard deviation of signal over b0 images, and the average SNR was obtained by averaging voxel-specific SNR values over the brain volume.

**Simulations**

The noise robustness and rotational variance of µFA estimation using the minimal design was studied by generating synthetic data by analytically simulating axially symmetric diffusion tensors of varying sizes and shapes. Synthetic signal was generated according to

$$S = S_0 e^{-b:D} \qquad (9)$$

where $S_0$ is the MRI signal without diffusion-weighting, **b** is the measurement tensor, **D** is the diffusion tensor, and : stands for the generalized tensor product[20]. The DDE sequence with parallel and orthogonal wave vectors encode linear and planar measurement tensors, respectively. Diffusion tensors were used instead of restricted diffusion because the minimal design assumes that diffusion can be approximated as Gaussian in every microscopic compartment[26]. The effects of non-Gaussian diffusion are outside the scope of this work. µFA was calculated from single-shell DDE data according to Equations 3 and 4 with MD equaling its ground truth value. The following simulations were performed.

**Simulation 1:** The rotational variance of the minimal design was quantified by simulating noise-free DDE experiments of single diffusion tensors with 400 unique orientations uniformly distributed over half a sphere. Both the 5-design and the minimal design were used with a total *b*-value of 2250 s/mm$_2$. The standard deviations of the resulting µFA estimates were calculated to estimate the rotational variances of the two protocols. µFA of the simulated tensors was varied from 0 to 1 and MD was varied from 0.1 to 3 µm$_2$/ms. 50 values for both µFA and MD were used leading to a total number of 10$_6$ different simulated tensors.

**Simulation 2:** The noise robustness of the minimal design was quantified by adding Rician noise to synthetic data and quantifying its effect on the estimate of µFA. Rician noise was emulated by adding real and imaginary Gaussian noise to the data and by taking the modulus of the resulting complex number to be the noisy signal. Here, SNR refers to the mean signal at $b = 0$ divided by the standard deviation of noise in each channel. Single tensors aligned with the x axis were used. µFA of the simulated tensors was varied from 0 to 1 and MD was varied from 0.1 to 3 µm$_2$/ms. 100 values for both parameters were used leading to a total number of 10$_4$ different simulated tensors. Simulations were performed using total *b*-values of 1500, 2250, 3000, and 3750 s/mm$_2$. Biophysically meaningless values of µFA, which are imaginary or greater than 1 due to noise, were rejected. Two noise propagation simulation experiments were performed. First, the standard deviations of $10^4$ noisy µFA estimates at three levels of SNR (10, 25, and 100) were calculated over the whole parameter space. Second, the minimum SNR required for measuring µFA within 0.1 from its asymptotic value with 95 % confidence was estimated by repeating the noisy simulations $10^3$ times over the whole parameter space at 100 levels of SNR equally distributed between 1 and 1000. The asymptotic value of the µFA estimate refers to the µFA value calculated from single-shell data with infinite SNR, which is a biased estimate of µFA[35].

## RESULTS

**Imaging experiments**

**Experiment 1:** The results of the first *ex vivo* imaging experiments are illustrated in Figures 2 and 3. Figure 2 shows the powder averaged signals for parallel and orthogonal wave vector orientations from a representative slice and at different *b*-values. SNR of the pre-processed data was already very high, typically around 150 for the b0 images, and the ensuing powder averaged signals are correspondingly robust for all *b*-values. The subtraction of these powder-averaged signals, dependent on µA, shows a *b*-value dependent pattern increasing in intensity with higher *b*-values, especially in white matter.

The µFA maps derived from Jespersen's DDE 5-design and Yang et al.'s DDE minimal design are shown in Figure 3A and 3B. Qualitatively, the maps are nearly indistinguishable, and when the maps are subtracted, no clear spatial pattern is observed (Figure 3C). When the µFA derived from each method were plotted against each other voxel-wise (Figure 3D), a very strong and statistically significant correlation was observed (Pearson's correlation coefficient = 0.91, $p < 10^{-10}$) with the data points very close to the unity line. More importantly, only a small bias was observed as the distribution of voxel-wise differences was centered near zero at 0.014 with a standard deviation of 0.07. Figure 3E shows the voxel-wise differences plotted against their mean value, revealing larger differences at lower values of µFA, as expected based on the analysis described in the theory section.

To test whether the small variance observed was due to the different methods or due to noise, we performed a simple test-retest experiment for the DDE 5-design (Figure 3F and 3G). The difference image again shows no particular spatial pattern (Figure 3H), and when the test-retest metrics were plotted against each other, a very similar pattern was observed than with the comparison of the minimal design against the 5-design (Figure 3I). Indeed, the variance in this test-retest experiment was also very similar to that observed in Experiment 1 above, with the distribution of voxel-wise differences between test and retest exhibiting a mean of 0.004 and a standard deviation of 0.06. The Pearson's correlation coefficient was 0.93 for the µFA maps acquired with two repetitions of the 5-design ($p < 10^{-10}$), and the data points closely followed the unity line. The voxel-wise differences are plotted against their mean values in Figure 3J, again revealing larger differences at lower values of µFA and showing that the precision of µFA decreases with decreasing microscopic anisotropy.

**Experiment 2:** In the second *ex vivo* experiment, the methods were compared over a larger interval of *b*-values which was more densely sampled to see if the results deviate at higher *b*-values due to differences in how non-Gaussian diffusion affects the powder averaged data acquired with the two methods. Results shown in Figure 4 confirmed that the accuracy of the minimal design is comparable to the 5-design with *b*-values up to 5000 s/mm$^2$. Figure 4 also shows that increasing diffusion-weighting leads to underestimation of µFA, because the maps were calculated without a higher order correction.

**Experiment 3:** Finally, we confirmed that our *ex vivo* experiments are representative of *in vivo* experiments by replicating the results *in vivo*, which can be seen in Figures 3K - 3O. The very similar µFA maps extracted from data acquired with the 5-design and the minimal design are shown in Figure 3K and 3L. The µFA values extracted from the two methods are highly correlated (Pearson's R = 0.86, p < 10$^{-10}$) with the data points following the unity line, and the voxel-wise difference distribution's mean is equal to 0.012 with a standard deviation of 0.10.

**Simulations**

**Simulation 1:** The standard deviations of µFA estimates derived from simulated measurements with different tensor orientations are shown in Figure 5. These results reveal that the minimal design is more rotationally variant than the 5-design when used for measuring µFA in individual tensors with high anisotropy or diffusivity. The standard deviation of µFA estimates ranges from 0 to 0.1 for the 5-design and 0 to 0.26 for the minimal design with the mean of the standard deviation of µFA estimates over the whole parameter space being 0.03 and 0.06 for the 5-design and the minimal design, respectively. No differences were observed in the mean values of the μFA estimates of tensors with different orientations obtained with the two protocols.

**Simulation 2:** Figure 6 portrays the standard deviation of μFA estimates from synthetic experiments repeated with random noise addition. The simulation results reveal that precision of μFA is highly dependent on its value and that particularly low values of MD, 0.5 μm$_2$/ms or less (depending on the SNR and the *b*-value), prevent precise measurement of μFA irrespective of its value. Greater diffusion-weighting increases the precision of the noisy estimate of small compartments' μFA but without enabling precise measurement of low values of μFA.

Figure 7 shows a map of the estimated minimum required SNR for measuring μFA within 0.1 margin from its asymptotic value with 95 % confidence with the minimal design. These results suggest that an SNR of 50 would enable reliable measurement of μFA values greater than 0.75, and that an SNR of nearly 200 would be necessary for reliably measuring μFA values of less than 0.5. The results also show that by increasing the magnitude of diffusion-weighting, reliable quantification of microscopic anisotropy becomes possible in smaller compartments. On the other hand, higher level of diffusion-weighting prevents measurements in large compartments due to the signal having hit the noise floor. These results suggest that to maximize precision in human brain tissue, where MD is usually around 0.8 μm$_2$/ms, it is desirable to use a *b*-value near 2250 s/mm$_2$.

# DISCUSSION

The DDE 5-design has been well established as a method capable of probing microscopic diffusion anisotropy, a clinically relevant microstructural property, in a model free fashion[19,23,35,38]. Yang et al.'s minimal design[26] enables crucial acceleration of the method for clinical applications, but its accuracy requires validation against the "gold standard" 5-design. Furthermore, to assign confidence in the precision of the µFA estimates given an SNR and maximal $b$-values, a noise propagation analysis is required and would be generally relevant both for clinical and preclinical applications. Hence, this study had two main purposes: (1) to experimentally validate the minimal design proposed by Yang et al. and (2) to assess µFA's noise robustness and to provide SNR requirements for reporting specific µFA values.

While Jespersen et al.'s DDE 5-design[19] – the contemporary "gold standard" for quantifying rotationally-invariant microscopic diffusion anisotropy[38] – requires at least 72 separate acquisitions per $b$-value, Yang's approach entails a minimum of only 12 acquisitions[26]. In our study, we applied the bipolar version of the minimal design, which consists of 24 acquisitions, due to potential cross-term effects[42]. After correcting for multiplexing effects (72 5-design acquisitions vs. 24 minimal design acquisitions averaged 3 times), a comparison of µFA maps acquired with each method revealed qualitatively indistinguishable results. In fact, µFA's variance in the test-retest of 5-design was very similar to the comparison of 5-design vs. minimal design. Thus, the powder averaged µFA calculated from 72 measurements of the DDE 5-design can be closely approximated in neural tissue, including both gray matter and white matter tissues, by Yang et al.'s minimal design[26] in the $b$-value range that was studied here. In addition, we have shown that the experimental validation holds both for *ex vivo* and *in vivo* conditions, thereby excluding tissue fixation effects as confounders. For every orthogonal gradient pair in the 5-design, there are three other gradient pairs encoding the same measurement tensor[19]. Thus, in case of Gaussian diffusion and infinite SNR, 15 orthogonal wave vector pairs are sufficient for measuring the same powder averaged signal than what is measured with the 5-design. In voxels with sufficient microscopic orientation dispersion, this can be further reduced to six as shown by Yang et al[26]. However, if the microscopic compartments are perfectly aligned, the minimal design suffers from larger rotational variance than the 5-design, as shown by the simulations presented here (Figure 5), which can lead to significantly biased results as previously discussed by Szczepankiewiczc et al[39].

Recently, the accuracy of μFA estimated by powder averaging methods such as the 5-design has been scrutinized, and Ianus et al. found significant higher order effects that bias the estimates at different *b*-values[35]. Corrections of the bias can be achieved by measuring more *b*-value shells and fitting the higher order terms[35,36]. However, every shell consumes more experimental time that is limited in clinical applications. In this study, we chose to forego higher order correction due to the prolonged acquisition; hence the μFA maps are biased due to *b*-value dependency. Still, it is worth noting that if the biased μFA is precise, it can be highly useful both in basic research and in the clinic. Our investigation into μFA's precision revealed two important features: μFA precision will depend strongly and nonlinearly on MD and on the actual μFA value. When MD is very small, the estimation of μFA is hampered even with very high SNR, and for reasonable values of MD, precise estimates of μFA < 0.5 require very high (> 200) SNR. On the other hand, large values of MD result in very diminished signal, impeding precise measurements. Our simulations also revealed the optimal *b*-value for probing micoscopic anisotropy in compartments with MD values of 0.8 $\mu m^2/ms$ to be near 2250 $s/mm^2$, a result which is consistent with previous studies[26]. Combined, these findings suggest that caution needs to be exercised in interpreting μFA maps, and the SNR should always be reported to assess the confidence in the estimates' precision. In particular, μFA maps are more reliable in areas with higher μFA, which should be taken into account upon interpretation of results. This lack of precision for low values of μFA have been previously reported with the μFA values derived from QTE experiments[18,39,47]. The large variance in voxels with low μFA can result in negative values of ε, leading to imaginary values of μFA. It is important to note that our simulations incorporated diffusion tensors, rather than restricted diffusion, whose effects were not the focus of this study. Still, given the experimental results, it seems that the Gaussian regime was not severely violated in our experiments or the µFA derived from the 5-design and minimal design would have differed.

Our results do not provide evidence that the minimal design is superior to the 5-design. Rather, the evidence points to an efficient acceleration which maps µFA approximately as robustly as the 5-design in neural tissue, when a fair comparison is performed. Therefore, given stringent time limitations that do not allow 72 acquisitions, microscopic diffusion anisotropy could be well-characterized with the approach proposed by Yang et al.[26]. Additionally, it may be possible to use the minimal design to shorten the acquisition time required for calculating µFA maps with the higher order correction[35,36], thereby providing more accurate metrics. However, three issues must be controlled for when applying the minimal design in the clinic. First, it is very challenging to reliably measure values of µFA lower than 0.5 (e.g., gray matter) with SNR typically achievable in the clinical setting, making it difficult to quantify microscopic diffusion anisotropy in gray matter without high gradients[48]. Second, the rotational variance of the minimal design may result in biased signal powder average in tissues where microenvironments are nearly aligned, highly anisotropic, and of large diffusivity[39]. Third, when using DDE with clinical scanners, it is important to control for concomitant fields, which can give rise to substantial signal bias[49]. Furthermore, we would like to point out that if the assumption of Gaussian diffusion is correct, then isotropic spherical tensor encoding with QTE would probably be even more suitable for mapping microscopic diffusion anisotropy in the clinical setting due to more efficient diffusion-weighting and larger signal deviation between the MDE and SDE acquisitions[20,50]. However, it is worth recalling that DDE offers model-free estimates while QTE's assumptions, as well as unresolved issues with time-dependent diffusion and rotational dependence[51], may incur other penalties in accuracy or precision of µFA estimates. DDE thus seems to offer a robust way for characterizing microscopic diffusion anisotropy.

# CONCLUSIONS

In conclusion, reducing the number of wave vector rotations in calculating the powder average of a DDE 5-design experiment, as proposed by Yang et al.[26], does not prevent the accurate quantification of microscopic diffusion anisotropy. The SNR requirements for precise quantification of μFA should be carefully considered when applying the discussed powder averaging scheme in a clinical setting for quantifying μFA.

# REFERENCES

1. Johansen-Berg H, Behrens TEJ. *Diffusion MRI*.; 2009. doi:10.1016/B978-0-12-374709-9.X0001-6
2. Basser PJ, Mattiello J, LeBihan D. MR diffusion tensor spectroscopy and imaging. *Biophys J*. 1994. doi:10.1016/S0006-3495(94)80775-1
3. Assaf Y, Johansen-Berg H, de Schotten M. The role of diffusion MRI in neuroscience. *NMR Biomed*. 2017:e3762.
4. Le Bihan D, Mangin JF, Poupon C, et al. Diffusion tensor imaging: Concepts and applications. *J Magn Reson Imaging*. 2001. doi:10.1002/jmri.1076
5. Sundgren PC, Dong Q, Gómez-Hassan D, Mukherji SK, Maly P, Welsh R. Diffusion tensor imaging of the brain: Review of clinical applications. *Neuroradiology*. 2004. doi:10.1007/s00234-003-1114-x
6. Lerner A, Mogensen MA, Kim PE, Shiroishi MS, Hwang DH, Law M. Clinical Applications of Diffusion Tensor Imaging. *World Neurosurg*. 2014. doi:10.1016/j.wneu.2013.07.083
7. Basser PJ, Pierpaoli C. Microstructural and physiological features of tissues elucidated by quantitative-diffusion-tensor MRI. *J Magn Reson - Ser B*. 1996. doi:10.1006/jmrb.1996.0086
8. Jones DK, Knösche TR, Turner R. White matter integrity, fiber count, and other fallacies: the do's and don'ts of diffusion MRI. *Neuroimage*. 2013. doi:10.1016/j.neuroimage.2012.06.081
9. Warach S, Chien D, Li W, Ronthal M, Edelman RR. Fast magnetic resonance diffusion-weighted imaging of acute human stroke. *Neurology*. 2012. doi:10.1212/wnl.42.9.1717
10. Sorensen AG, Buonanno FS, Gonzalez RG, et al. Hyperacute stroke: evaluation with combined multisection diffusion-weighted and hemodynamically weighted echo-planar MR imaging. *Radiology*. 2014. doi:10.1148/radiology.199.2.8668784
11. Clark CA, Barrick TR, Murphy MM, Bell BA. White matter fiber tracking in patients with space-occupying lesions of the brain: A new technique for neurosurgical planning? *Neuroimage*. 2003. doi:10.1016/j.neuroimage.2003.07.022
12. Jensen JH, Helpern JA, Ramani A, Lu H, Kaczynski K. Diffusional kurtosis imaging: The quantification of non-Gaussian water diffusion by means of magnetic resonance imaging. *Magn Reson Med*. 2005. doi:10.1002/mrm.20508
13. Kiselev VG. The Cumulant Expansion: An Overarching Mathematical Framework For Understanding Diffusion NMR. In: *Diffusion MRI*. ; 2013. doi:10.1093/med/9780195369779.003.0010
14. Callaghan PT, MacGowan D, Packer KJ, Zelaya FO. High-resolution q-space imaging in porous structures. *J Magn Reson*. 1990. doi:10.1016/0022-2364(90)90376-K
15. Callaghan PT, Eccles CD, Xia Y. NMR microscopy of dynamic displacements: K-space and q-space imaging. *J Phys E*. 1988. doi:10.1088/0022-3735/21/8/017
16. Wedeen VJ, Hagmann P, Tseng WYI, Reese TG, Weisskoff RM. Mapping complex tissue architecture with diffusion spectrum magnetic resonance imaging. *Magn Reson Med*. 2005. doi:10.1002/mrm.20642
17. Shemesh N, Jespersen SN, Alexander DC, et al. Conventions and nomenclature for double diffusion encoding NMR and MRI. *Magn Reson Med*. 2016. doi:10.1002/mrm.25901
18. Lasič S, Szczepankiewicz F, Eriksson S, Nilsson M, Topgaard D. Microanisotropy imaging: quantification of microscopic diffusion anisotropy and orientational order parameter by diffusion MRI with magic-angle spinning of the q-vector. *Front Phys*. 2014;2:11. doi:10.3389/fphy.2014.00011
19. Jespersen SN, Lundell H, Sønderby CK, Dyrby TB. Orientationally invariant metrics of apparent compartment eccentricity from double pulsed field gradient diffusion experiments. *NMR Biomed*. 2013. doi:10.1002/nbm.2999

20. Topgaard D. Multidimensional diffusion MRI. *J Magn Reson*. 2017. doi:10.1016/j.jmr.2016.12.007
21. Eriksson S, Lasic S, Topgaard D. Isotropic diffusion weighting in PGSE NMR by magic-angle spinning of the q-vector. *J Magn Reson*. 2013. doi:10.1016/j.jmr.2012.10.015
22. Westin CF, Knutsson H, Pasternak O, et al. Q-space trajectory imaging for multidimensional diffusion MRI of the human brain. *Neuroimage*. 2016. doi:10.1016/j.neuroimage.2016.02.039
23. Özarslan E. Compartment shape anisotropy (CSA) revealed by double pulsed field gradient MR. *J Magn Reson*. 2009. doi:10.1016/j.jmr.2009.04.002
24. Lasič S, Szczepankiewicz F, Eriksson S, Nilsson M, Topgaard D. Microanisotropy imaging: quantification of microscopic diffusion anisotropy and orientational order parameter by diffusion MRI with magic-angle spinning of the q-vector. *Front Phys*. 2014;2:11. doi:10.3389/fphy.2014.00011
25. Szczepankiewicz F, van Westen D, Englund E, et al. The link between diffusion MRI and tumor heterogeneity: Mapping cell eccentricity and density by diffusional variance decomposition (DIVIDE). *Neuroimage*. 2016. doi:10.1016/j.neuroimage.2016.07.038
26. Yang G, Tian Q, Leuze C, Wintermark M, McNab JA. Double diffusion encoding MRI for the clinic. *Magn Reson Med*. 2018. doi:10.1002/mrm.27043
27. Jespersen SN. Equivalence of double and single wave vector diffusion contrast at low diffusion weighting. *NMR Biomed*. 2012;25(6):813-818.
28. Jespersen SN, Buhl N. The displacement correlation tensor: microstructure, ensemble anisotropy and curving fibers. *J Magn Reson*. 2011;208(1):34-43.
29. Cory DG, Garroway AN, Miller JB. Applications of spin transport as a probe of local geometry. In: *American Chemical Society, Polymer Preprints, Division of Polymer Chemistry*. ; 1990.
30. Shemesh N, Özarslan E, Komlosh ME, Basser PJ, Cohen Y. From single-pulsed field gradient to double-pulsed field gradient MR: Gleaning new microstructural information and developing new forms of contrast in MRI. *NMR Biomed*. 2010. doi:10.1002/nbm.1550
31. Mitra PP. Multiple wave-vector extensions of the NMR pulsed-field-gradient spin-echo diffusion measurement. *Phys Rev B*. 1995. doi:10.1103/PhysRevB.51.15074
32. Shemesh N, Özarslan E, Adiri T, Basser PJ, Cohen Y. Noninvasive bipolar double-pulsed-field-gradient NMR reveals signatures for pore size and shape in polydisperse, randomly oriented, inhomogeneous porous media. *J Chem Phys*. 2010. doi:10.1063/1.3454131
33. Shemesh N, Cohen Y. Microscopic and compartment shape anisotropies in gray and white matter revealed by angular bipolar double-PFG MR. *Magn Reson Med*. 2011. doi:10.1002/mrm.22738
34. Lawrenz M, Koch MA, Finsterbusch J. A tensor model and measures of microscopic anisotropy for double-wave-vector diffusion-weighting experiments with long mixing times. *J Magn Reson*. 2010. doi:10.1016/j.jmr.2009.09.015
35. Ianuş A, Jespersen SN, Serradas Duarte T, Alexander DC, Drobnjak I, Shemesh N. Accurate estimation of microscopic diffusion anisotropy and its time dependence in the mouse brain. *Neuroimage*. 2018. doi:10.1016/j.neuroimage.2018.08.034
36. Shemesh N. Axon Diameters and Myelin Content Modulate Microscopic Fractional Anisotropy at Short Diffusion Times in Fixed Rat Spinal Cord. *Front Phys*. 2018. doi:10.3389/fphy.2018.00049
37. Lampinen B, Szczepankiewicz F, Mårtensson J, van Westen D, Sundgren PC, Nilsson M. Neurite density imaging versus imaging of microscopic anisotropy in diffusion MRI: A model comparison using spherical tensor encoding. *Neuroimage*. 2017. doi:10.1016/j.neuroimage.2016.11.053
38. Henriques RN, Jespersen SN, Shemesh N. Microscopic anisotropy misestimation in spherical-mean single diffusion encoding MRI. *Magn Reson Med*. 2019. doi:10.1002/mrm.27606
39. Szczepankiewicz F, Sjölund J, Ståhlberg F, Lätt J, Nilsson M. Tensor-valued diffusion encoding for diffusional variance decomposition (DIVIDE): Technical feasibility in clinical MRI systems. *PLoS One*. 2019;14(3):1-20. doi:10.1371/journal.pone.0214238

40. Jespersen SN, Lundell H, Sønderby CK, Dyrby TB. Commentary on "Microanisotropy imaging: quantification of microscopic diffusion anisotropy and orientation of order parameter by diffusion MRI with magic-angle spinning of the q-vector." *Front Phys*. 2014;2:28.
41. Ku HH. Notes on the use of propagation of error formulas. *J Res Natl Bur Stand Sect C Eng Instrum*. 2012. doi:10.6028/jres.070c.025
42. Neeman M, Freyer JP, Sillerud LO. A simple method for obtaining cross-term-free images for diffusion anisotropy studies in NMR microimaging. *Magn Reson Med*. 1991. doi:10.1002/mrm.1910210117
43. Veraart J, Novikov DS, Christiaens D, Ades-Aron B, Sijbers J, Fieremans E. Denoising of diffusion MRI using random matrix theory. *Neuroimage*. 2016. doi:10.1016/j.neuroimage.2016.08.016
44. Kellner E, Dhital B, Kiselev VG, Reisert M. Gibbs-ringing artifact removal based on local subvoxel-shifts. *Magn Reson Med*. 2016. doi:10.1002/mrm.26054
45. Guizar-Sicairos M, Thurman ST, Fienup JR. Efficient subpixel image registration algorithms. *Opt Lett*. 2008.
46. Garyfallidis E, Brett M, Amirbekian B, et al. Dipy, a library for the analysis of diffusion MRI data. *Front Neuroinform*. 2014. doi:10.3389/fninf.2014.00008
47. Szczepankiewicz F, Lasič S, van Westen D, et al. Quantification of microscopic diffusion anisotropy disentangles effects of orientation dispersion from microstructure: Applications in healthy volunteers and in brain tumors. *Neuroimage*. 2015. doi:10.1016/j.neuroimage.2014.09.057
48. Lawrenz M, Finsterbusch J. Detection of microscopic diffusion anisotropy in human cortical gray matter in vivo with double diffusion encoding. *Magn Reson Med*. 2019;81(2):1296-1306. doi:10.1002/mrm.27451
49. Szczepankiewicz F, Westin CF, Nilsson M. Maxwell-compensated design of asymmetric gradient waveforms for tensor-valued diffusion encoding. *Magn Reson Med*. 2019. doi:10.1002/mrm.27828
50. Sjölund J, Szczepankiewicz F, Nilsson M, Topgaard D, Westin C-F, Knutsson H. Constrained optimization of gradient waveforms for generalized diffusion encoding. *J Magn Reson*. 2015;261:157-168.
51. Jespersen SN, Olesen JL, Ianuş A, Shemesh N. Effects of nongaussian diffusion on "isotropic diffusion" measurements: An ex-vivo microimaging and simulation study. *J Magn Reson*. 2019. doi:10.1016/j.jmr.2019.01.007

# FIGURES

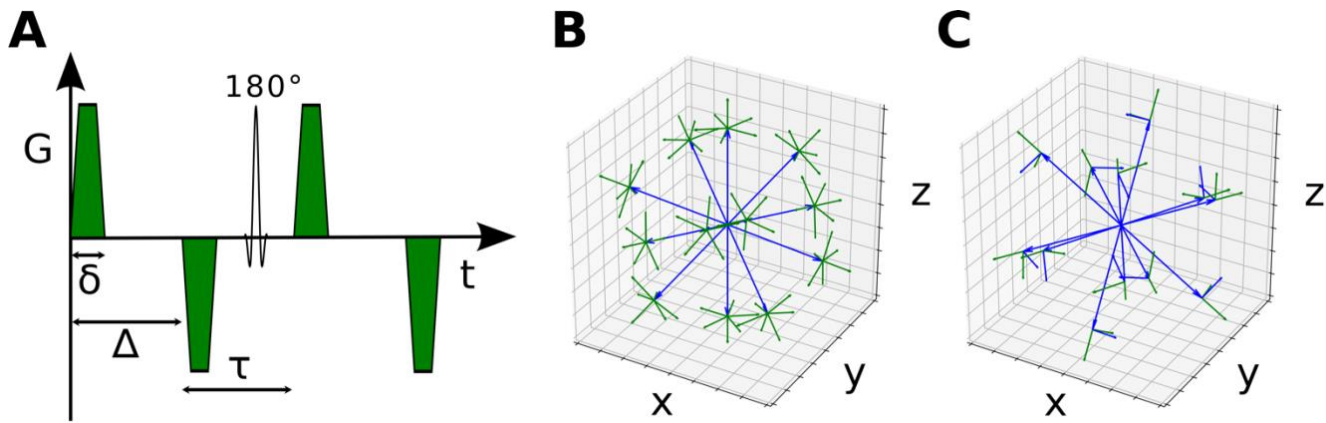

**Figure 1.** A) A schematic representation of the used DDE sequence where gradient pulse pairs of equal magnitude are separated by a 180° RF pulse. $\Delta$ is the diffusion time, $\delta$ is the pulse duration, and $\tau$ is the mixing time. B) Gradient directions of the 5-design. C) Gradient directions of the recently proposed minimal design. In B and C, the directions of the first wave vector are shown in blue and the directions of the second wave vector are shown in green. The first wave vectors pointing towards the vertices of the icosahedron are represented by longer arrows.

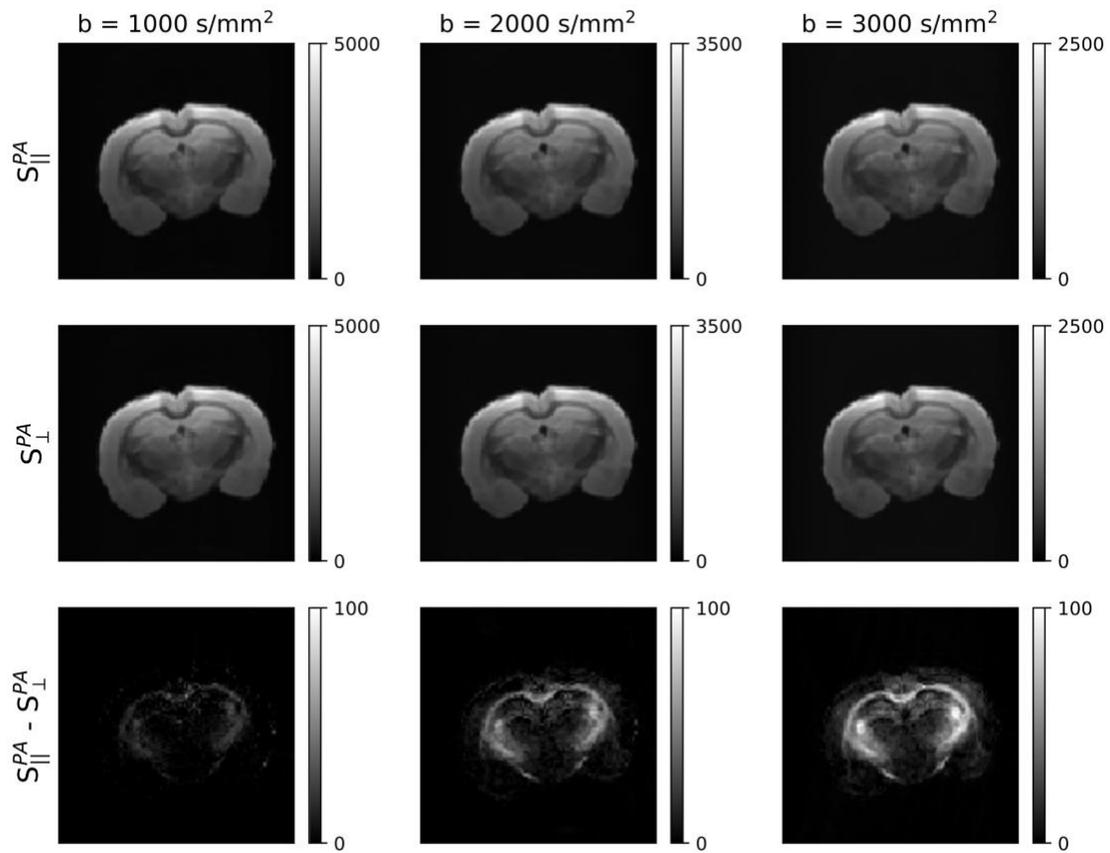

**Figure 2.** Powder averaged data from experiment 1 acquired with parallel (first row) and orthogonal wave vectors (second row) with three b-values. Microscopic anisotropy can be observed by subtracting the powder averaged data acquired with orthogonal wave vectors from the powder averaged data acquired with parallel wave vectors (third row).

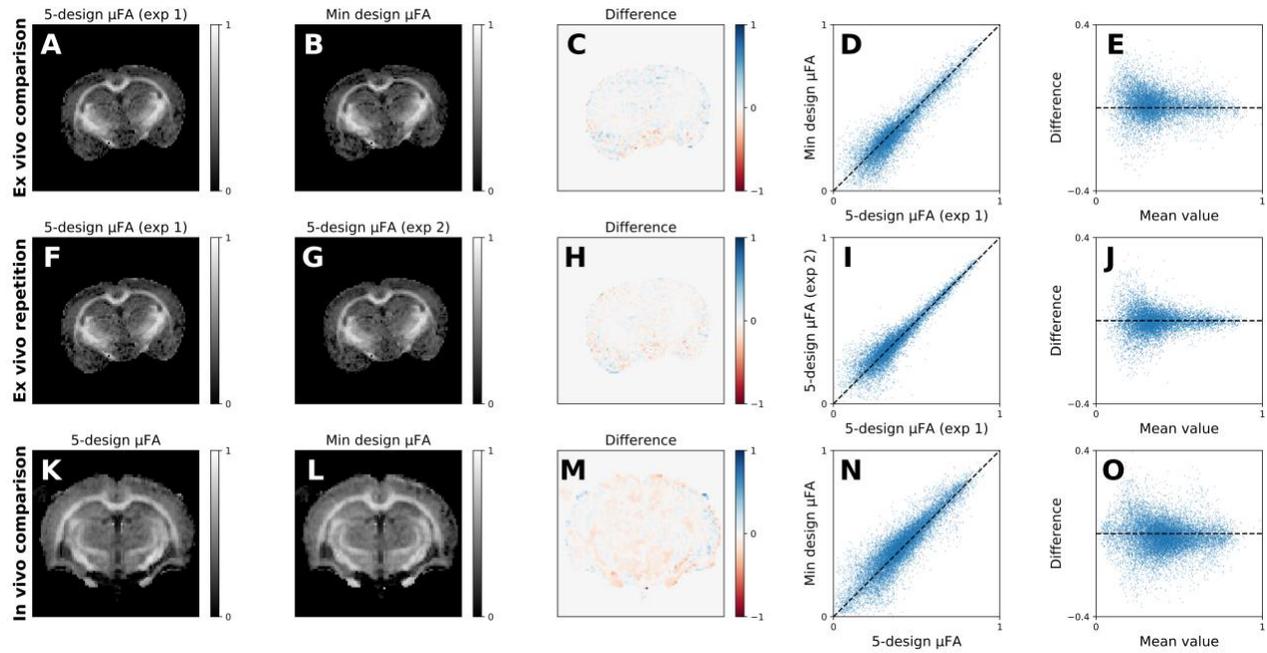

**Figure 3.** Comparison of μFA maps calculated with the 5-design and the minimal design. Top row shows a comparison between the 5-design and the minimal design (Experiment 1). Middle row shows the same comparison between two repetitions of the 5-design experiment (Experiment 1). Bottom row shows the results of the in vivo experiment (Experiment 3). Difference stands for the map obtained by reducing the map in the second column from the map in the first column. The fourth and fifth columns show the voxel-wise comparison of the maps. The dashed lines in the fourth column are identity lines.

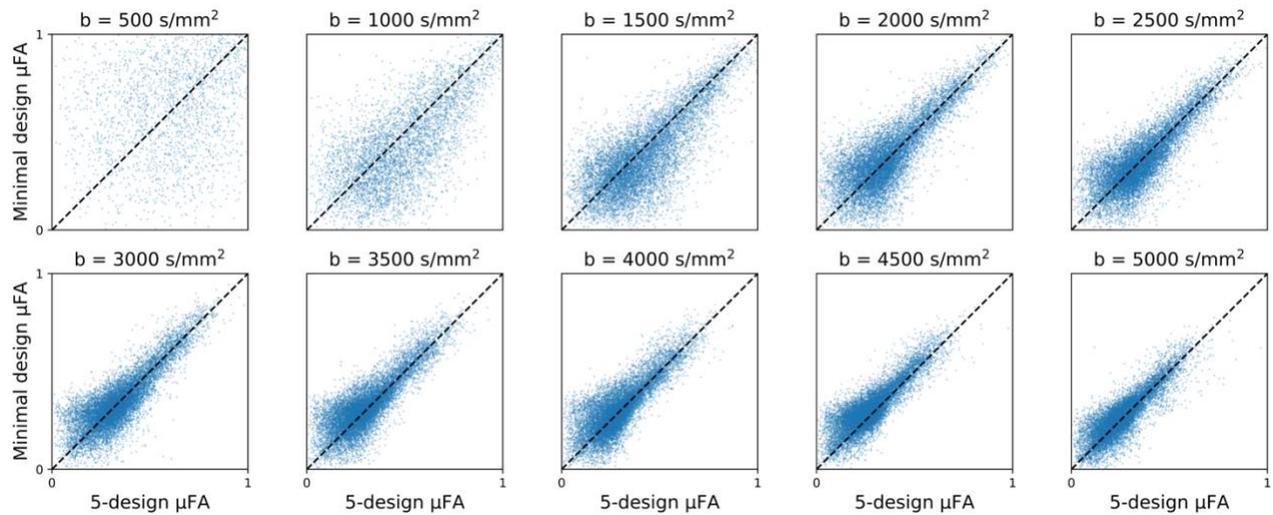

**Figure 4.** Voxel-wise comparison of µFA maps calculated from single-shell data (experiment 2) acquired with the 5-design and the minimal design over a range of b-values. The dashed lines are identity lines.

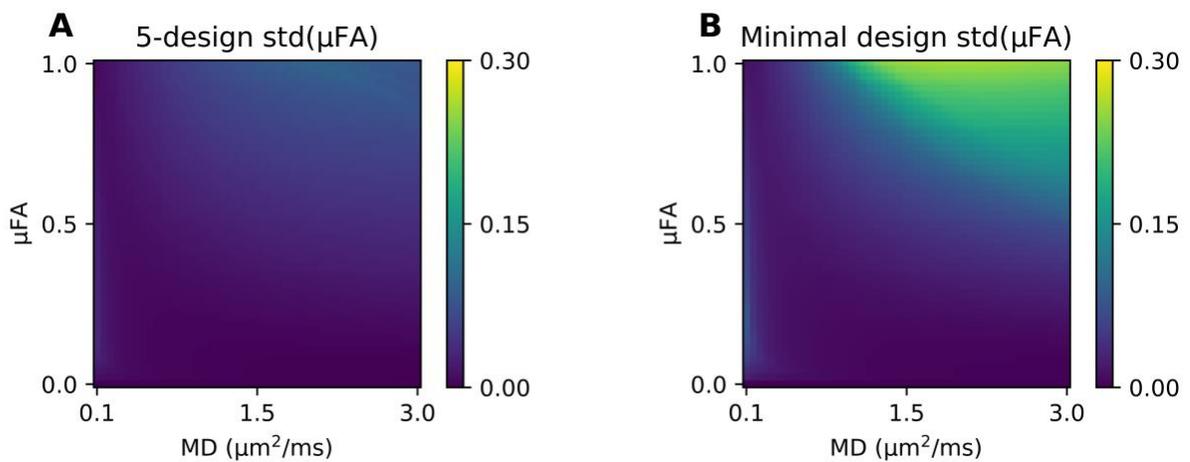

**Figure 5.** Standard deviations of µFA estimates from simulated experiments of 400 axially symmetric tensors with directions uniformly distributed over half a sphere with the 5-design (A) and the minimal design (B) using a b-value of 2250 s/mm$_2$.

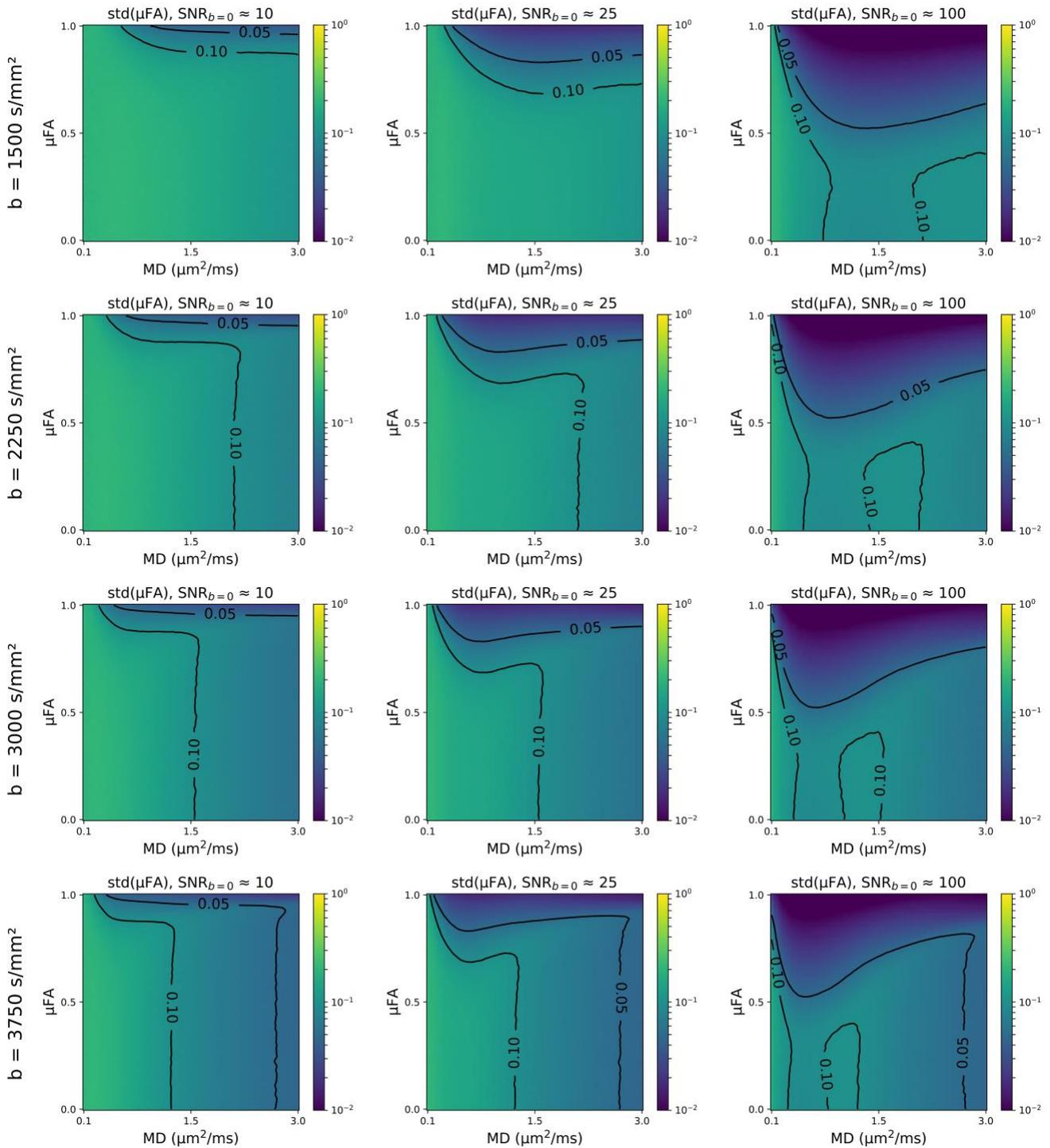

**Figure 6.** The standard deviations of μFA estimates calculated from 10⁴ repetitions of simulations with three different levels of noise and with four b-values. Simulations were performed with single diffusion tensors with varying anisotropies and mean diffusivities. Biophysically meaningless values of μFA were excluded from calculations. The contour lines follow data smoothed with a Gaussian filter ($\sigma = 0.8$ pixels).

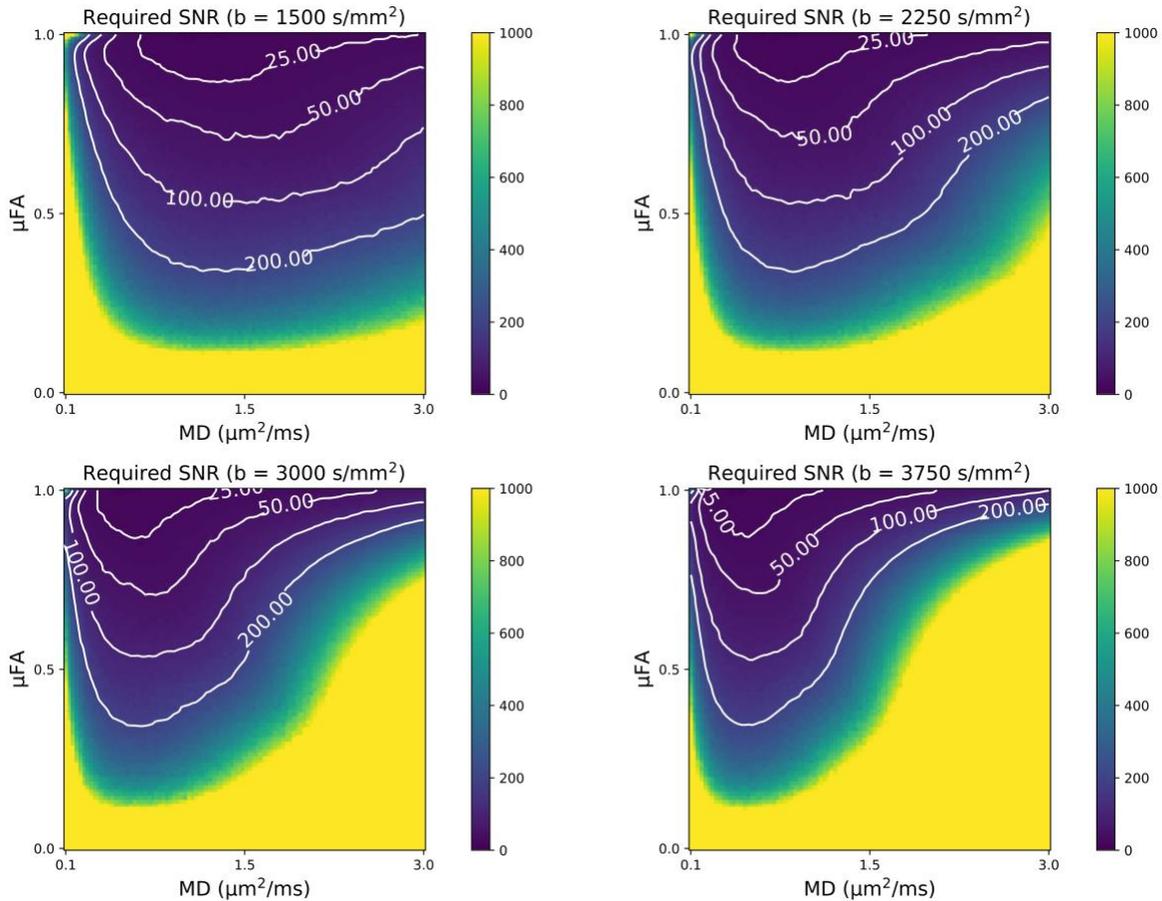

**Figure 7.** The minimum SNR required for measuring µFA within 0.1 margin from its asymptotic value with 95 % confidence with four different b-values and varying diffusion tensor sizes and shapes. The figure was generated from $10^3$ repetitions of noisy simulations at 100 levels of SNR equally distributed between 1 and 1000. Tensors requiring a higher SNR than 1000 are colored as yellow. The contour lines follow data smoothed with a Gaussian filter ($\sigma = 0.8$ pixels).